# About Quartz Crystal Resonator Noise: Recent Study

F. Sthal[a], S. Galliou[a], J. Imbaud[a], X. Vacheret[a], P Salzenstein[a], E. Rubiola[a] and G. Cibiel[b]

[a]*Time and Frequency Dept., FEMTO-ST Institute, UMR CNRS 6174, ENSMM, UFC, UTBM ENSMM, 26 Chemin de l'Epitaphe, 25030 Besançon cedex, France*
[b]*Microwave and Time-Frequency Department, CNES, Toulouse, France*

**Abstract.** The first step, before investigating physical origins of noise in resonators, is to investigate correlations between external measurement parameters and the resonator noise. Tests and measurements are mainly performed on an advanced phase noise measurement system, recently set up. The resonator noise is examined as a function of the sensitivity to the drive level, the temperature operating point and the tuning capacitor.

## INTRODUCTION

The Centre National d'Etudes Spatiales (CNES), Toulouse, France and FEMTO-ST Institute, Besancon, France, have initiated a program of investigations on the origins of noise in bulk acoustic wave resonators. Several European manufacturers of high quality resonators and oscillators are involved in this operation [1]. Tests and measurements are mainly performed on an advanced phase noise measurement system, recently set up for this program [2-3]. The instrument sensitivity, that is, the background phase noise converted into Allan deviation, is of $10^{-14}$. Understanding quartz crystal resonator noise is a complex problem and requires a long-term study. This will be performed in two main steps.

The first step is to investigate correlations between actual geometrical data, external parameters of influence and the measured resonator parameters. The second step will consist in building a noise model for bulk acoustic wave resonators from a more microscopic basis. The inherent noise of a quartz crystal resonator is now clearly identified as 1/f noise and it is the main limitation of ultra-stable quartz crystal oscillators [4, 5]. For this study, according to a macroscopic point of view, three research axes have been investigated: resonators sensitivity to the drive level, effect of temperature (influence of the operating point versus the turn over temperature) and the frequency shift (pull-up). First results of the analyses are presented in this paper. Several sets of measurements have been compared.

# NOISE CHARACTERIZATION METHOD

Carrier suppression technique is used in the bench principle. The general idea of this passive method (Fig. 1a) consists in reducing the noise of the source as much as possible [4-5]. Indeed, when resonators exhibit a very weak noise, the noise of the source is always higher than that of the quartz crystal resonator. Thus, the direct feeding of the driving source signal through only one resonator does not permit to extract the resonator noise from the output resulting noise. On the other hand, the source signal can be subtracted when passing through two identical arms equipped with identical resonators (the devices under tests: DUT). Then the contribution of the source is cancelled while inner noise of both resonators is preserved because one resonator noise is not correlated to the other one. When the carrier suppression is achieved (less than -75 dBc is acceptable), the resulting signal only made up noise from both resonators, is strongly amplified and mixed with the source signal to be shifted down to the low frequency domain and processed by the spectrum analyzer. In such a way, noise to be measured from both resonators can be brought up at a higher level than the driving source noise. Moreover, the noise floor of the bench can be measured with resistors substituted for crystal resonators. Fig. 1b shows the measurement system. Boxes concept gives a lot of facilities to build the system.

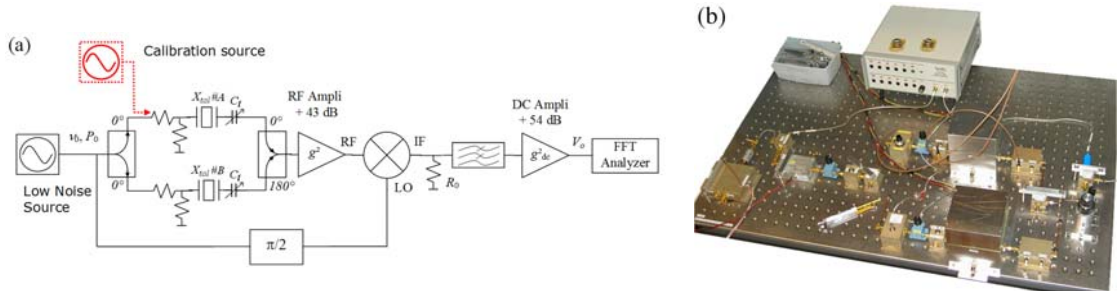

**FIGURE 1.** (a) Principle of the measurement bench, (b) Resonator phase noise measurement bench.

Fig. 2 gives a typical result of the measured single side-band power spectral density of the phase fluctuations, $\mathcal{L}(f) = S_\phi(f)/2$. The floor of the short-term stability of the resonator is classically given by the Allan standard deviation. Considering the 1/f resonator frequency fluctuation of the resonator, the measured value at 1 Hz of the power spectral density of phase fluctuations, $S_\phi(1\,Hz)$, is used to give the flicker floor due to the resonator.

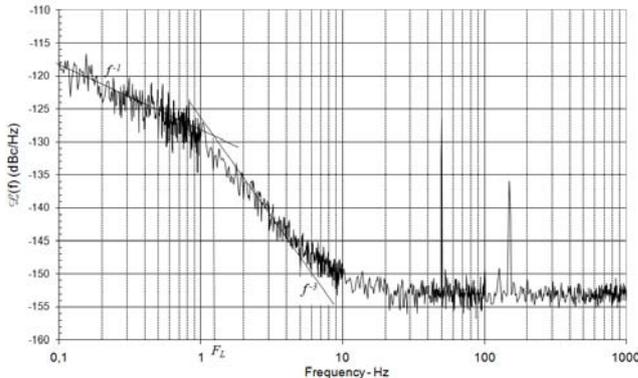

$$\sigma_{y\_floor} = \frac{F_L}{f_{res.}}\sqrt{2\ln 2\, S_\phi(1Hz)}$$

with $F_L$ the Leeson frequency and $f_{res}$ the resonance frequency of the resonator.

**FIGURE 2.** Typical $\mathcal{L}(f)$ of 5 MHz SC-cut resonator. In this case $\sigma_{y\text{–floor}} = 1\cdot 10^{-13}$.

# EXPERIMENTAL RESULTS

## Drive Level effect

Measurements have been carried out on 10 MHz, SC-cut, quartz crystal resonator pairs from manufacturer A and 5 MHz SC-cut resonators from manufacturer B. As the resonator dissipated power in oscillator is about 100 µW, the measurement power is varied to 20, 100 and 200 µW. The tuning series capacitor $C_t$ remains constant when power is changed. The frequency shift induced by the amplitude-frequency effect is compensated by the output frequency of the synthesizer. Fig. 3a shows the Allan standard deviation versus the crystal dissipated power $P_{xtal}$. The trend of the curves is similar in regard to the 10 or 5 MHz resonators. Fig. 3b shows $\sigma_{yfloor}$ according to the crystal dissipated power density. The crystal power density is computed considering the acoustic volume of the vibration [6]. Influence of this parameter on the noise level is not demonstrated.

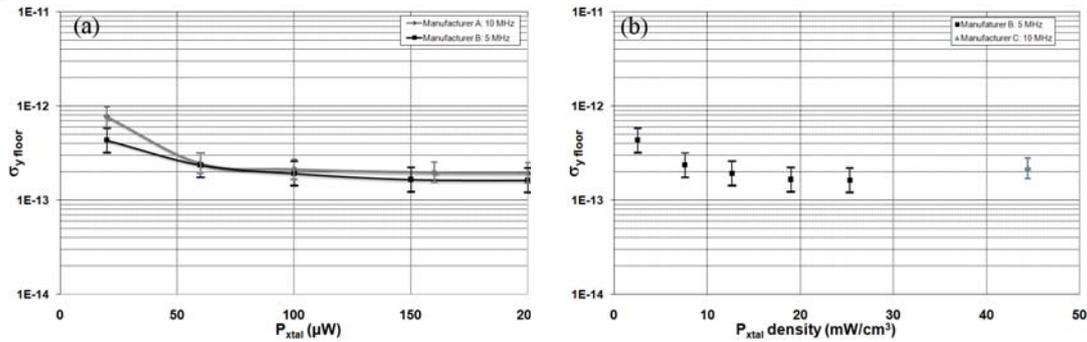

**FIGURE 3.** (a) $\sigma_{y\_floor}$ vs. crystal dissipated power, (b) $\sigma_{y\_floor}$ vs. crystal power density. For the 10 MHz resonators, $C_t$ is equal to 60 pF and for the 5 MHz, $C_t$ is equal to 33 pF.

## Effect of Temperature

The influence of the temperature is particularly studied according to the operating point ($T_{op}$) versus the quartz crystal turn over temperature ($T_{to}$). Fig. 4a shows the frequency temperature behaviour of the quartz crystal resonator. Thermally controlled ovens have been used in order to control the quartz crystal temperature by step lower than 0.05°C [7]. Noise measurements have been done on classical 10 MHz, SC-cut quartz crystal resonator. Fig. 4b shows that the modification of the temperature operating point has not a real influence on the resonator noise.

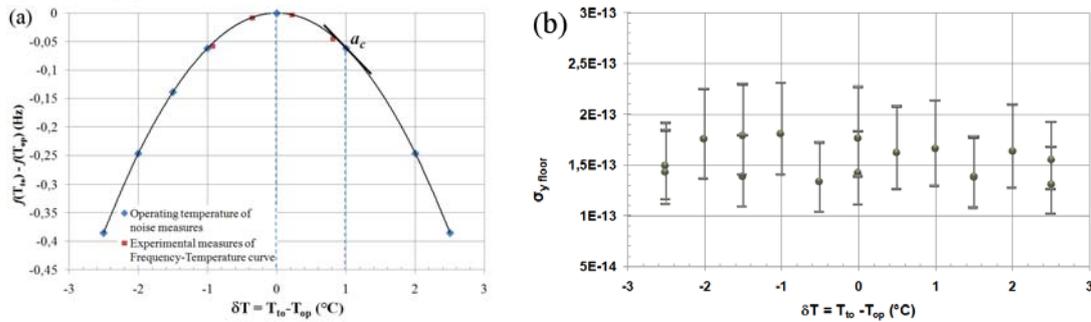

**FIGURE 4.** (a) Frequency variation vs. temperature of a SC cut resonator, (b) Allan standard deviation $\sigma_{y\_floor}$ of a resonator pair measured versus the temperature shift $\delta T$.

## Pull-Up Capacitor

Fig. 5a illustrates the frequency shift induced by the serial capacitor. In Fig. 5b Allan standard deviation of a 10 MHz, SC cut resonator is given according to $C_t$ when capacitor values have been chosen at 10 pF, 60 pF and 220 pF. All noise values are inside the error bars except for the smallest value of $C_t$. This simple modelization shows that a small value of the tuning capacitor $C_t$ should not be chosen. The best configuration is given according to the parallel static capacitor of the resonator [3].

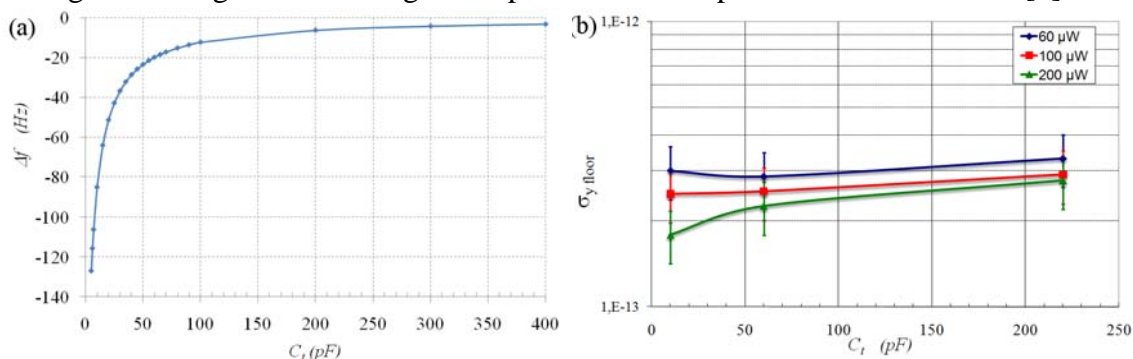

**FIGURE 5.** (a) Frequency shift versus $C_t$ of a 10 MHz, SC-cut resonator. (b) Allan standard deviation, $\sigma_{y\_floor}$, according to $C_t$.

## CONCLUSION

Several sets of measurements have been compared. The resonator noise is observed according to the input power in the bench arm. About the temperature influence, a stable oven allows to reject the relative frequency fluctuations due to the oven stability at a lower level than the quartz crystal noise. Because of the measurement errors, the resonator noise seems not influenced by the resonator operating temperature. The resonator noise is also measured at a constant resonator power according to a wide range of series capacitors. The tuning capacitor modifies the overall impedance, but, a correlation between load impedance, tuning capacitor and the flicker noise of the resonator is not shown. The second step will now consist in building a noise model for bulk acoustic wave resonators from a more microscopic basis. Such a conference is a unique opportunity for us to compare our work with that of the noise community and to deal with noise physicists.

## REFERENCES

1. S. Galliou *et al*, Proc. Joint Meeting IEEE Ann. Freq. Cont. Symp. and European Freq. and Time Forum, Genova, Switzerland, 2007, pp. 1176-1181.
2. F. Sthal *et al*, Proc. Joint Meeting IEEE Ann. Freq. Cont. Symp. and European Freq. and Time Forum, Genova, Switzerland, 2007, pp. 254-260.
3. F. Sthal *et al*, Proc. 20th European Freq. and Time Forum, Toulouse, France, 2008, FPE-0057.pdf.
4. F. Sthal *et al*, IEEE Trans. Ultrason. Ferroelec. Freq. Contr., vol. 47, 369-373, (2000).
5. E. Rubiola *et al*, IEEE Trans. Ultrason. Ferroelec. Freq. Contr., vol. 47, 361-368, (2000).
6. D.S. Stevens *et al*, J. A. S. A., vol 79, 6, 1811-1826, (1986).
7. F. Sthal *et al*, Electronics Letters, vol. 43, no. 16, 900-901, (2007).